\documentclass[aps,prl,twocolumn,a4paper,superscriptaddress,twoside,showpacs,showkeys]{revtex4}
\usepackage{amsmath,amssymb,amsfonts}
\usepackage{graphicx}
\usepackage[english]{babel}
\usepackage{psfrag}
\usepackage{color}
\usepackage{tabularx}
\usepackage{times}

\newcommand{\Sb}{\mathbf{S}}
\newcommand{\Jp}{$J$-$J^\prime$}
\renewcommand{\onlinecite}{\cite}

\begin{document}

\title{Evidence of Unconventional Universality Class in a
  Two-Dimensional Dimerized Quantum Heisenberg Model} \author{Sandro
  Wenzel} \email{wenzel@itp.uni-leipzig.de} \affiliation{Institut
  f\"ur Theoretische Physik and Centre for Theoretical Sciences (NTZ), Universit\"at Leipzig, Postfach 100 920,
  D-04009 Leipzig, Germany} \author{Leszek Bogacz}
\affiliation{Department of Information Technology, Jagiellonian
  University, ul. Reymonta 4, 30-059 Krakow, Poland} \author{Wolfhard
  Janke} \email{janke@itp.uni-leipzig.de} \affiliation{Institut
  f\"ur Theoretische Physik and Centre for Theoretical Sciences (NTZ), Universit\"at Leipzig, Postfach 100 920,
  D-04009 Leipzig, Germany}
\date{\today}

\begin{abstract}
The two-dimensional $J$-$J^\prime$ dimerized quantum Heisenberg model is
studied on the square lattice by means of (stochastic series
expansion) quantum Monte Carlo simulations as a function of the
coupling ratio \hbox{$\alpha=J^\prime/J$}. The critical point of the
order-disorder quantum phase transition in the $J$-$J^\prime$ model is
determined as \hbox{$\alpha_\mathrm{c}=2.5196(2)$} by finite-size
scaling for up to approximately $10\,000$ quantum spins. By comparing six
dimerized models we show, contrary to the current belief, that the
critical exponents of the $J$-$J^\prime$ model are not in agreement with the
three-dimensional classical Heisenberg universality class. This lends
support to the notion of nontrivial critical excitations at the
quantum critical point.
\end{abstract}


\maketitle
Dimerized quantum spin systems are important examples of
low-dimensional antiferromagnets featuring a quantum phase transition
(QPT) \cite{sachdev:qpt} which destroys a N\'eel ordered state by
competition between different interactions. In contrast to other
examples showing such criticality, in this class of models the actual
transition is triggered by nonisotropic couplings where the dimers
\footnote{Or quadrumers, etc.} are explicitly placed on the lattice.
Because of the discovery of Bose-Einstein condensation of magnons in a
magnetic field much effort has been spent to study their physics
\cite{giamarchi-2008-4}.

The characteristics of the QPT in two-dimensional (2D) dimerized
models have been investigated in detail. By mapping to a nonlinear
sigma model (NLSM) \cite{chakravarty_prl} it was argued that the
transition is well described by the Heisenberg O(3) classical
universality class in three dimensions (3D). The role of Berry phase
terms, which are present in the mapping to the NLSM, is argued to be
irrelevant \cite{chubukov_prl} and there are numerous numerical
studies which support this claim. Examples include the CaVO lattice
\cite{troyer-1997-66}, bilayer models \cite{wang:014431} and the 2D
coupled ladder system \cite{PhysRevB.65.014407}.

Recently, the idea of deconfined quantum critical points has been put
forward by Senthil \emph{et al.}  \cite{senthil-2004-303} who argue
that there are, however, also important examples of QPTs where Berry
phases and nontrivial excitations at the quantum critical point can
change the critical behavior. These arguments are based on Heisenberg
models with \emph{isotropic} interactions exhibiting a transition
between two ordered states such as an antiferromagnetic and a
valence-bond solid phase, and numerical evidence for such a case was
recently claimed by Sandvik \cite{sandvik-2007}.  This idea that
challenges the standard Landau-Ginzburg-Wilson framework of phase
transitions was also found to be relevant in other systems such as
classical dimers \cite{alet-2006-97} in 3D and has prompted further
theoretical and numerical efforts
\cite{kragset-2006-97,nogueira-2007-76,motrunich_2004,nazario_2006,wiese_2008},
some of which extend to different scenarios or show that the field is
still highly controversial.

In this context, it is in any case somewhat surprising that also a
specific 2D dimerized spin model, which we refer to as the \Jp{}
model, with \emph{nonisotropic} interactions was suggested as a
candidate for deconfinement at the quantum critical point by Yoshuika
\emph{et al.}  \cite{yoshioka:174407} (see also
Ref.~\cite{takashima-2006-73}). In consequence this idea could lead to
critical exponents characterizing the phase transition, which differ
from those of the Heisenberg universality class in 3D. This conclusion
was, however, questioned \cite{senthil-2005-74} because of the close
relation of the \Jp{} model to the ladder model.
In order to resolve this conflict and to give arguments in favour of
one or the other alternative we report in this Letter on quantum Monte
Carlo (QMC) simulations of various dimerized models, which signal the
emergence of an unconventional phase transition for the \Jp{} model.
\begin{figure}[b]
\begin{minipage}{0.49\columnwidth}
\psfrag{l}{}
\psfrag{j1}{{\color{blue} $J^\prime$}}
\psfrag{j2}{{\color{red} $J$}}
  \includegraphics[width=0.95\textwidth]{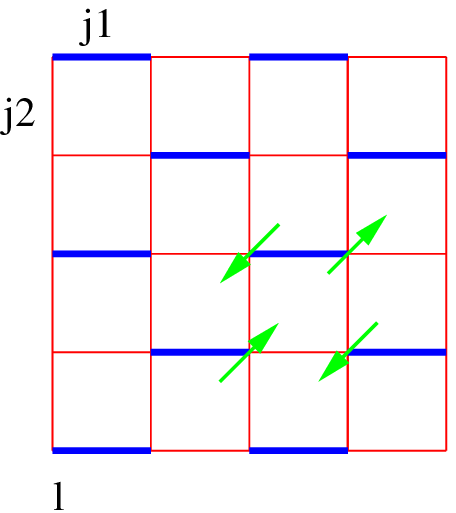}\\[-0.5cm]
\hspace{-3.1cm}(a)
\end{minipage}
\begin{minipage}{0.49\columnwidth}
\psfrag{l}{}
\psfrag{j1}{{\color{blue} $J^\prime$}}
\psfrag{j2}{{\color{red} $J$}}
\includegraphics[width=0.95\textwidth]{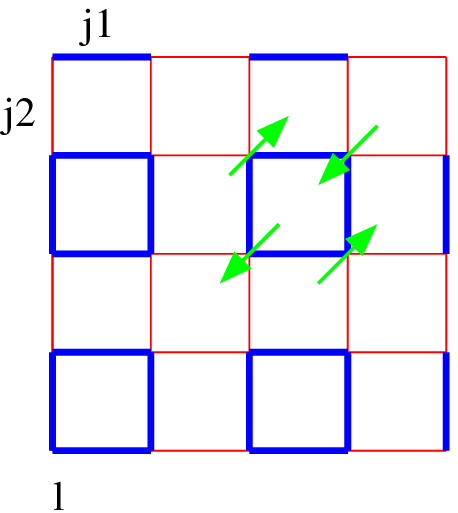}\\[-0.5cm]
\hspace{-3.1cm}(b)
\end{minipage}
\caption{\label{fig:jjp}(color online). (a) Visualization of the \Jp{} model on the 2D square lattice. The quantum spin ($S=1/2$) degrees of freedom
  live on a square lattice with different nearest neighbor
  couplings $J$ and $J^\prime$ (thin and thick). (b) Similar for the plaquette model, favoring quadrumer formation.}
\end{figure}
\begin{figure*}
\begin{minipage}{0.33\textwidth}
\includegraphics[width=\textwidth]{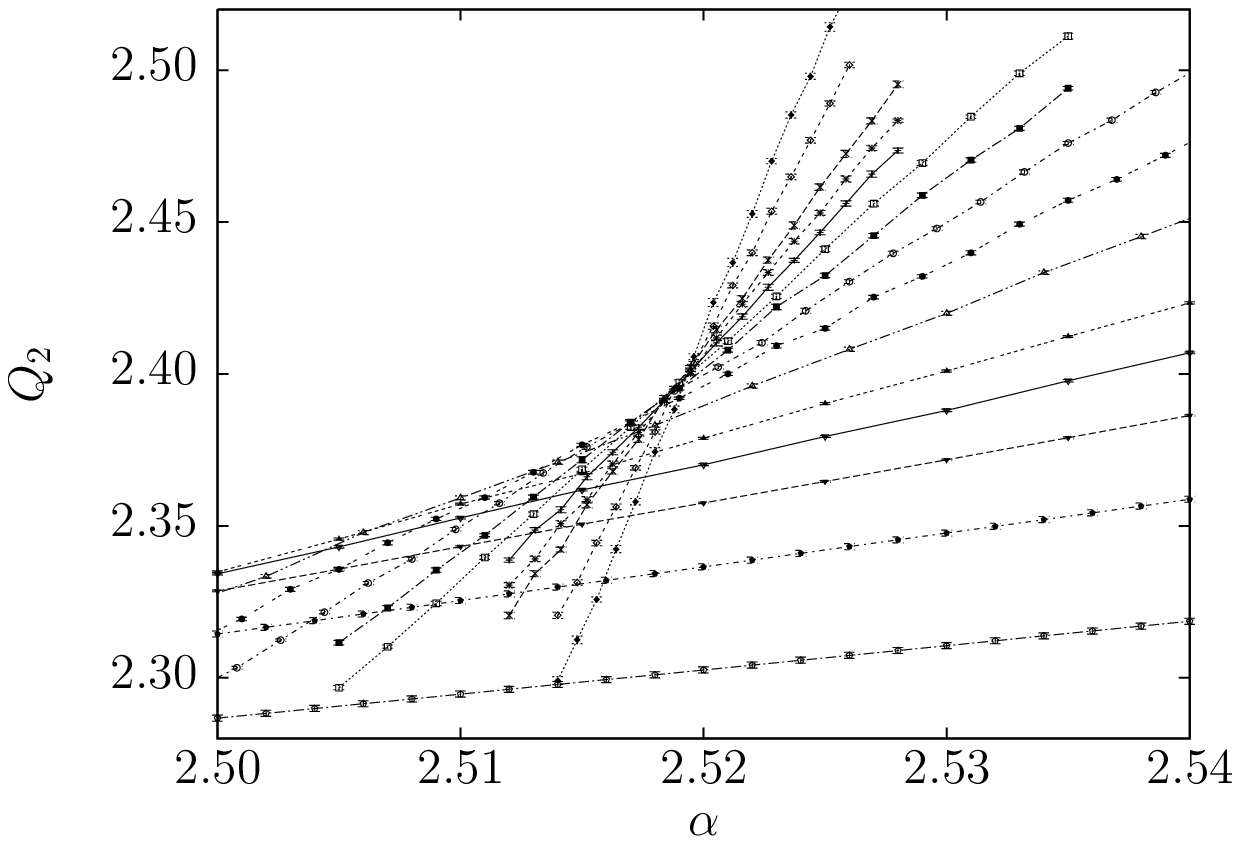}
\end{minipage}
\begin{minipage}{0.33\textwidth}
\includegraphics[width=\textwidth]{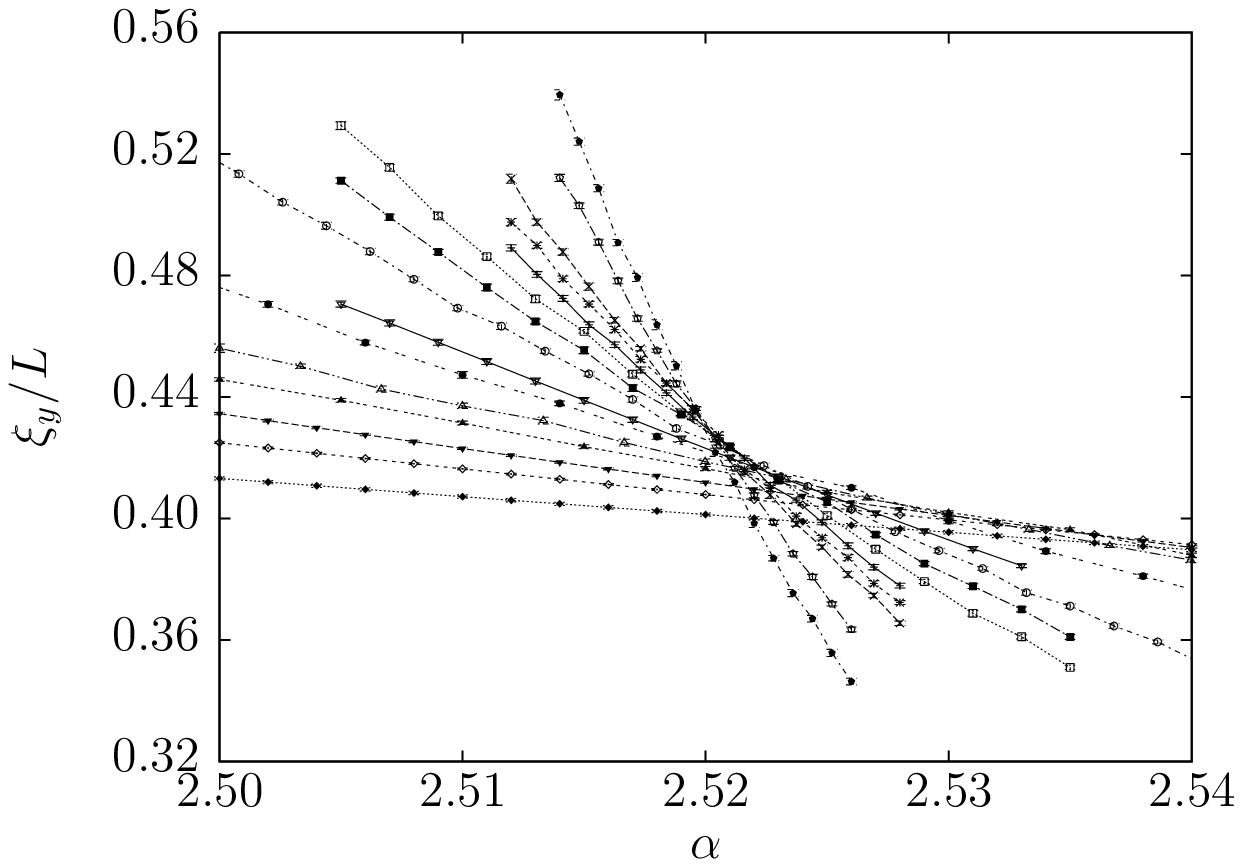}
\end{minipage}
\begin{minipage}{0.33\textwidth}
\includegraphics[width=\textwidth]{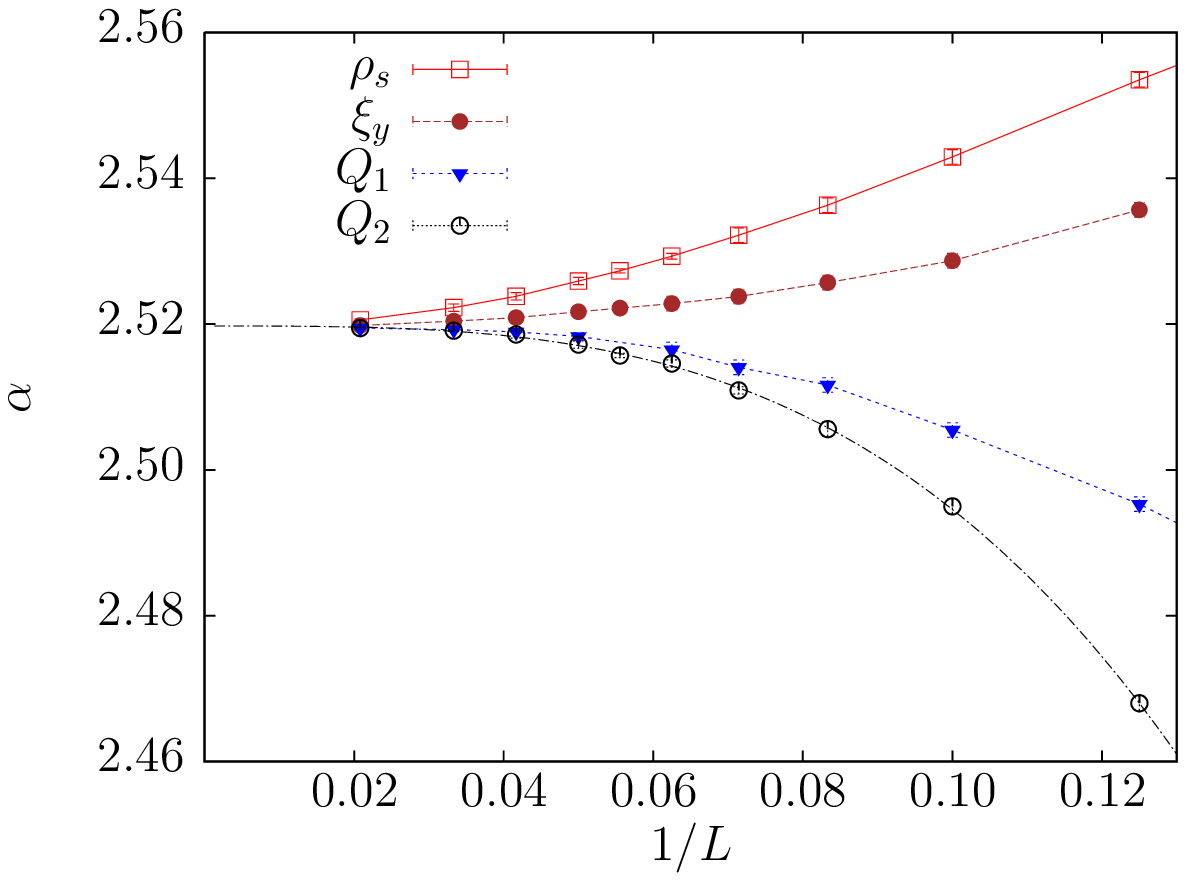}
\end{minipage}
\begin{picture}(0,0)
\put(-120,20){(a)}
\put(0,20){(b)}
\put(120,20){(c)}
\end{picture}
\caption{\label{fig:observables}(color online). (a) The Binder parameter $Q_2$ and (b) the
  correlation length $\xi_y/L$ for various lattice sizes from
  $L=8$ to $L=72$. (c) Scaling of the crossing points can be used to
  extract the critical coupling $\alpha_\mathrm{c}$. All quantities seem to converge to
  the same estimate.}
\end{figure*}

The \Jp{} model is defined on a square lattice with $N=L^2$ spins by the Hamiltonian
\begin{equation}
\label{egn:hamiltonian}
\mathcal{H}=J\sum_{\langle i,j\rangle}\Sb_i \Sb_j+J^\prime \sum_{\langle i,j\rangle^\prime}\Sb_i \Sb_j\,.
\end{equation}
Here, $\Sb_i=(1/2)\,(\sigma_x,\sigma_y,\sigma_z)$ denotes the usual
spin-$1/2$ operator at lattice site $i$, and $J$ and $J^\prime$ are
the antiferromagnetic coupling constants defined on the bonds $\langle
i,j \rangle$ and $\langle i,j \rangle^\prime$, respectively.  The
``staggered'' arrangements of the bonds on a square lattice with
periodic boundary conditions can be seen in Fig.~\ref{fig:jjp}(a).
The geometry of the ladder model results by a simple shift of every
second dimer. We define $\alpha=J^\prime/J$ as the parameter driving
the phase transition.

Simulations are performed with the directed loop variant
\cite{PhysRevE.66.046701} of the stochastic series expansion (SSE)
algorithm \cite{PhysRevB.43.5950} for lattice sizes $L=8$ up to $L=72$
(in single cases $L=96$) and inverse temperature up to $\beta=256$. We
checked that all quantities took on their ground-state values at the
temperature simulated and we scaled $\beta\sim L$. Additional parallel
tempering (PT) updates as well as multihistogram reweighting were
performed to further optimize sample statistics and data analysis.

To probe the nature of the quantum phase transition, we calculate
several well-known observables starting from the staggered
magnetization (the N\'eel order parameter) with
\begin{equation}
m^z_\mathrm{s}=\frac{1}{N}\sum_i^N S_i^z (-1)^{x_i+y_i}\,,
\end{equation}
and its Binder parameters $Q_1=\langle (m_\mathrm{s}^z)^2
\rangle/\langle |m_\mathrm{s}^z| \rangle^2$ and $Q_2= \langle
(m_\mathrm{s}^z)^4 \rangle / \langle (m_\mathrm{s}^z)^2 \rangle^2$.
These quantities are complemented by the second-moment correlation
length obtained from structure factors $S$ as
\begin{equation}
  \xi_y = \frac{L_y}{2\pi}\sqrt{\frac{S(\pi,\pi)}{S(\pi,\pi+2\pi/L_y)}-1}\,,
\end{equation}
with the obvious relation for the imaginary time correlation length
$\xi_\tau$. Lastly we determine the spin stiffness obtained from
\begin{equation} \rho_s=\frac{3}{4\beta N}\langle w_x^2 +
  w_y^2\rangle\,,
\end{equation}
where $w_x^2$ is the square of the difference of operator numbers
$S^+S^-$ and $S^-S^+$ in $x$-direction.  At a critical point the
quantities $Q_1$, $Q_2$, $\xi_y/L$ as well as $\rho_\mathrm{s}L$ are
expected to cross for different lattice sizes $L$ (under the
assumption that the imaginary time exponent $z=1$, in case of the spin
stiffness).

We first present QMC results for $Q_2$ and $\xi_y/L$ in
Fig.~\ref{fig:observables} where the crossing behavior becomes
evident. A second-order phase transition is therefore very likely to
happen. However, corrections to scaling terms are clearly present as
the crossing points are not sharp but rather spread out for smaller
lattice sizes. We exploit this fact by studying the scaling of the
crossing points at lattice sizes $L$ and $2L$ for the various
quantities. In this way a bracketing of the critical coupling
$\alpha_\mathrm{c}$ is obtained in Fig.~\ref{fig:observables}(c) and
we can easily read off a preliminary estimate as $\alpha_\mathrm{c}\in
[2.5190,2.5202 ]$. This value is made more precise by fitting to a
function
$\alpha_\mathrm{c}(L,2L)=\alpha_\mathrm{c}+aL^{-1/\nu-\omega}$
yielding $\alpha_\mathrm{c}=2.5198(3)$.  All observables agree in this
picture, indicating a single phase transition and our estimate is in
accordance to earlier quotes in the literature
\cite{PhysRevLett.61.2484,PhysRevB.53.2633,PhysRevB.61.14607,sandvik:207201}.

\paragraph{Finite-size scaling:}
Having gained a fairly good estimate of the critical coupling
$\alpha_\mathrm{c}$ we now turn to determining the critical exponent
$\nu$. This is done by using the scaling ansatz for a second-order
phase transition $\mathcal{O}_L(t)=L^{\lambda/\nu}g_\mathcal{O}
(tL^{1/\nu})$, where $\lambda$ is the scaling exponent associated with
the quantity $\mathcal{O}$, $t=\alpha/\alpha_\mathrm{c}-1$ the reduced
critical coupling and $g_\mathcal{O}$ a scaling function. For the
quantities $Q_2$ and $\xi_y/L$, it is clear that $\lambda=0$, which is
indeed \textit{verified} from the data in Fig.~\ref{fig:observables}.
In this work we follow Ref.~\onlinecite{wang:014431} and take evident
corrections to scaling explicitly into account by performing the data
analysis according to a more general scaling ansatz
\begin{equation}
\label{eqn:scaling}
\mathcal{O}_L(t)=L^{\lambda/\nu}(1 + cL^{-\omega})g_\mathcal{O} (tL^{1/\nu} + dL^{-\phi/\nu})\,,
\end{equation}
where $\omega$ is the usual confluent correction exponent and $\phi$ a
shift correction contribution. This way we can directly compare with a very
detailed study recently performed on two bilayer models
favoring dimer formation which gave strong support
for O(3) universality \cite{wang:014431}.%
\begin{figure}
\includegraphics[width=0.85\columnwidth]{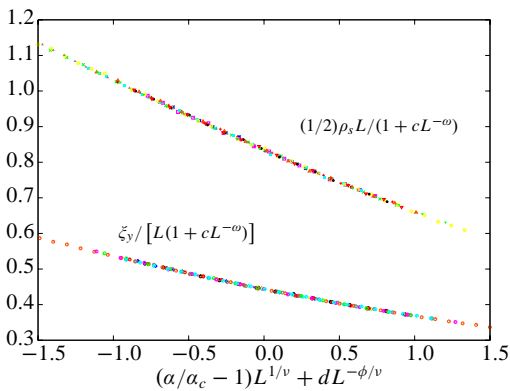}
\caption{\label{fig:collapse}(color online). Best collapse for the spin stiffness (upper data set) and the correlation length (lower data set) obtained from fitting data to the scaling ansatz \eqref{eqn:scaling}.}
\end{figure}
We perform our data analysis using this scaling ansatz in two ways.
First, a Taylor expansion of $g_\mathcal{O}(x)$ up to fourth order
in $x$ is used in conjunction with multidimensional fitting.
Second, we check this procedure by using a collapsing tool
\cite{wenzel:npb} which makes direct use of multi-histogram
reweighting. Both methods give consistent results for the critical
coupling ratio as $\alpha_\mathrm{c}=2.5196(2)$ and the critical
exponent $\nu=0.689(5)$ which is more than $4\sigma$ smaller than the
standard O(3) value \cite{hasenbuschO3} of $\nu=0.7112(5)$. We arrive
at this result conclusively for all observables of this study.  The
error bar reflects checking for different window sizes, as well as
trying different correction terms. In fact, we find that in most
cases the $\omega$ correction is sufficient, i.e., inclusion of $\phi$
terms does not change the estimate for $\nu$.
Figure~\ref{fig:collapse} contains a data collapse for all lattice
sizes for quantities $\xi_y$ and $\rho_\mathrm{s}L$.  The exponent $z$
can best be estimated from the correlation length $\xi_\tau$ in
imaginary time. As the result $\nu_\tau=0.687(5)$ is almost equal to
the previous value we can conclude $z=1.01(1)$.

\paragraph{Cross-checks and critical scaling:}
Since the discrepancy between $\nu$ and the standard O(3) value is
rather small we have performed multiple checks of the algorithm and
our numerical procedure in different categories. First, we repeat
the study for a different aspect ratio $2L_y \times L_y$ ($L_x=2L_y$)
(and larger $\beta$), where the correlation lengths in the $x$ and in
the $y$ direction are approximately equal, giving a consistent result
of $\nu=0.688(5)$.  Second, we carefully ran simulations on various
dimerized models known to be described by exponents in the 3D
Heisenberg universality class. These include the CaVO, the bilayer,
and the ladder model discussed before. In all cases we arrive easily
at Heisenberg universality. This is also true for the plaquette model
of Fig.~\ref{fig:jjp}(b), being a model not previously investigated to
high precision. There, we obtain a critical point of
$\alpha_\mathrm{c}=1.8228(4)$ and a critical exponent of
$\nu=0.709(8)$ using exactly the same procedure (even at smaller
lattice sizes of up to $L=48$) \cite{wenzel_jjp_prb}.
\begin{figure}
\begin{minipage}{\columnwidth}
\includegraphics[width=0.9\textwidth]{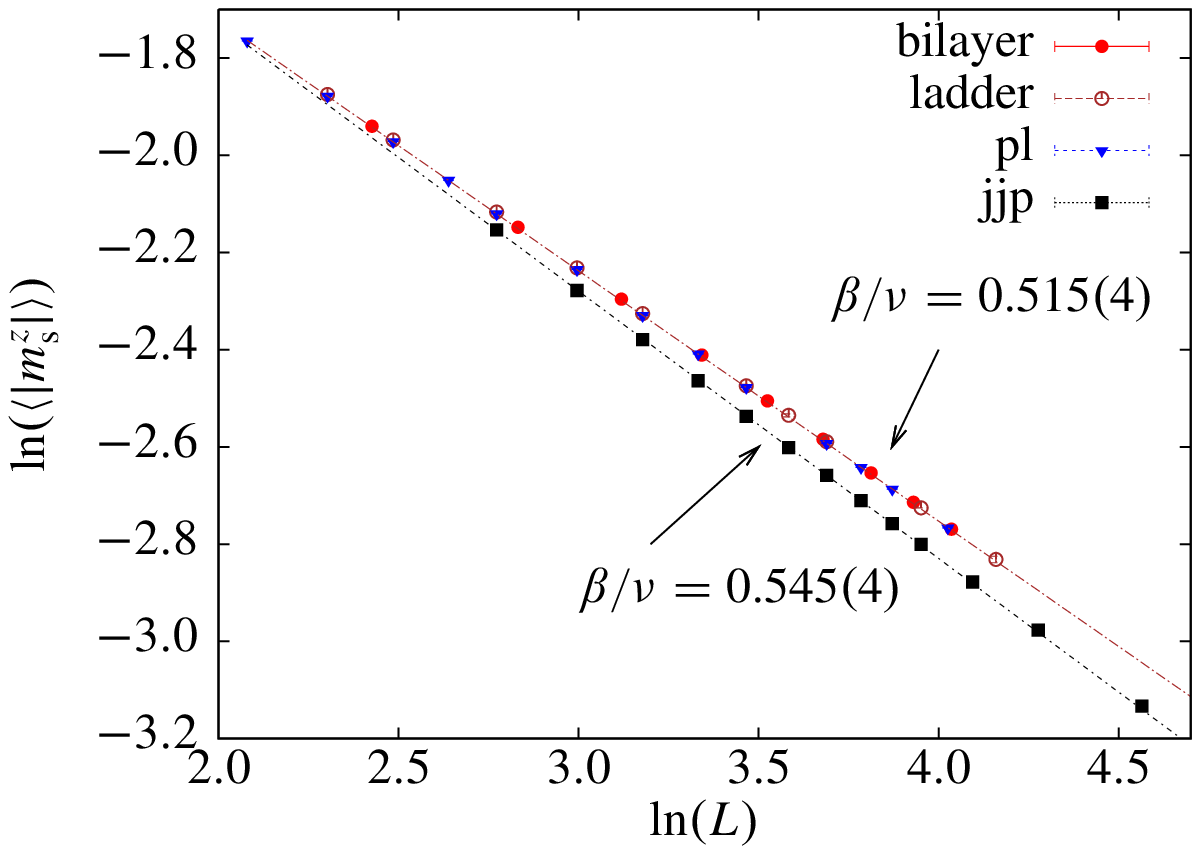}
\end{minipage}
\begin{minipage}{\columnwidth}
\includegraphics[width=0.9\textwidth]{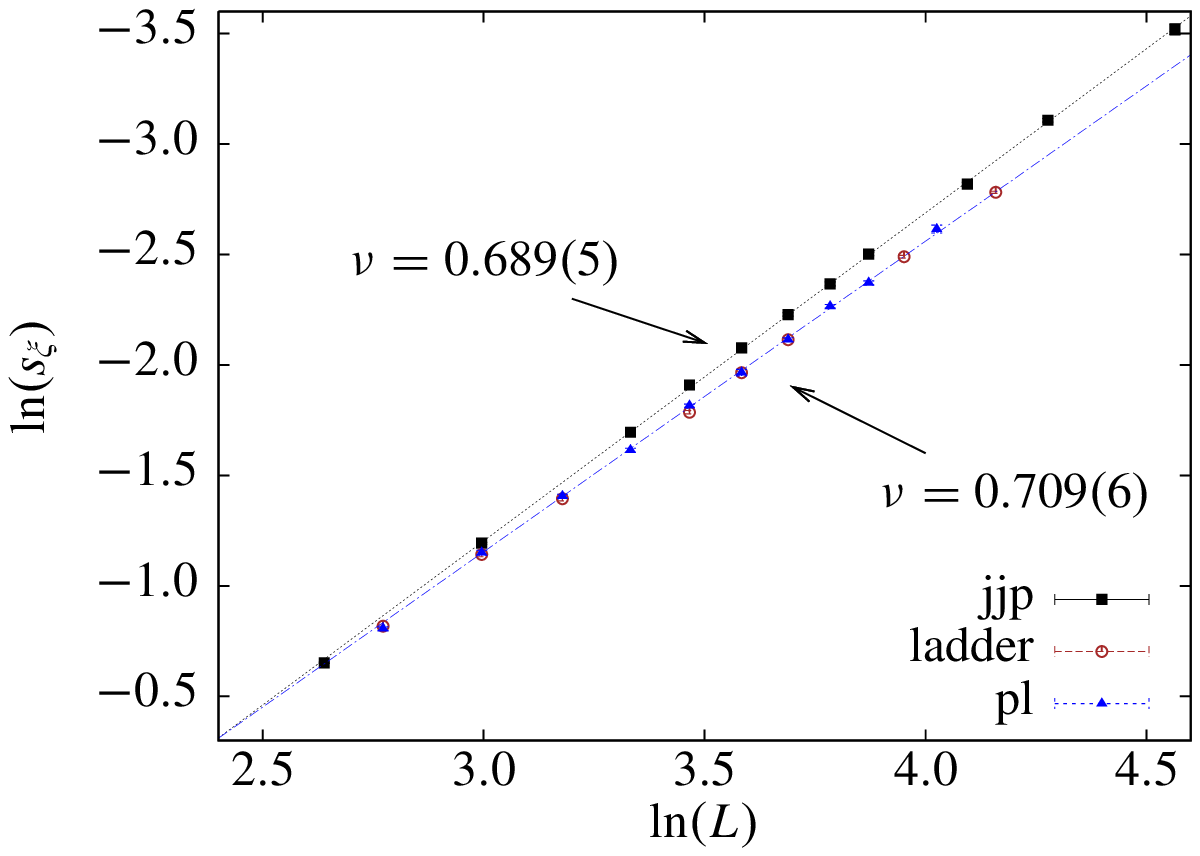}
\begin{picture}(0,0)
\put(-172,180){(a)}
\put(-172,130){(b)}
\end{picture}
\end{minipage}
\caption{\label{fig:smcmp} (color online). Scaling of (a) the staggered
  magnetization and the (b) the slope $s_\xi$ of $\xi_y/L$ at the
  critical point. We compare the scaling of the \Jp{} model (jjp) against
  the plaquette (pl), the bilayer and the ladder models at the best known
  critical couplings. Both quantities indicate different critical exponents
  for the \Jp{} model.}
\end{figure}
Those checks on known and hitherto less studied models indicate that
the critical exponent $\nu$ is indeed smaller for the \Jp{} model. 

To further investigate the ``mismatch'' of the universality class we
proceed with determining other critical exponents by studying the
scaling at the quantum critical point $\alpha_\mathrm{c}$. In this
case the staggered magnetization scales as $\langle | m_{\mathrm{s}}^z |\rangle \sim L^{-\beta/\nu}$ and we can obtain the exponent $\eta$ (as well as
$z$) from $\langle m_\mathrm{s}^2 \rangle$ and the staggered
susceptibility from
\begin{equation}
  \langle L^2 m_\mathrm{s}^2\rangle \sim L^{d-z-\eta}\,,\quad\quad\chi_\mathrm{s}\sim L^{\gamma/\nu}\,.
\end{equation}%
\begin{table}[b]
  \caption{\label{tab:cmpQ2} Heisenberg dimer and quadrumer systems used for comparison of scaling at the critical point.}
\renewcommand{\arraystretch}{1.2}
\begin{tabularx}{\columnwidth}{X X l l}
\hline\hline
Model & Type & $\alpha_\mathrm{c}$ & Reference \\
\hline
Bilayer & Symm., dimer & $2.5220(1)$ & \onlinecite{wang:014431}\\
Kondo  & Symm., dimer & $1.3888(1)$ & \onlinecite{wang:014431}\\
CaVO  & Symm., plaquette & $0.939(2)$ & \onlinecite{troyer-1997-66}\\
Plaquette  & Symm., plaquette & $1.8228(4)$  & \onlinecite{wenzel_jjp_prb} \\
Ladder & Unsymm., dimer & $1.909(1)$ & \onlinecite{PhysRevB.65.014407, wenzel_jjp_prb}\\
\Jp & Unsymm., dimer & $2.5196(2)$ & this work, \onlinecite{wenzel_jjp_prb} \\
\hline\hline
\end{tabularx}
\end{table}%
Using this approach, we obtain in Fig.~\ref{fig:smcmp}(a) the
estimates $\beta/\nu=0.515(4)$ for the bilayer, the ladder and the
plaquette model at the known critical points (see
Table~\ref{tab:cmpQ2}), and $\beta/\nu=0.545(4)$ for the \Jp{} model,
which should be compared to the O(3) value of $\beta/\nu=0.518(1)$
\cite{hasenbuschO3}. The error bars on the data reflect uncertainties
in $\alpha_\mathrm{c}$ and the straight line fits are all excellent
and results are independent of different fitting windows. Our results
are quoted for the five largest lattice sizes.  To make the
discrepancy in $\beta/\nu$ more apparent we have rescaled the original
data to start at a common point in the plot. Second, we compute
$\nu$ again from the slope $s_Q=\mathrm{d}Q_2/\mathrm{d\alpha}$ and
$s_\xi=(1/L)\mathrm{d}\xi_y/\mathrm{d\alpha}$ at the critical point
which should scale with lattice size as $L^{1/\nu}$.
Figure~\ref{fig:smcmp}(b) shows this for $s_\xi$ in comparison for the
\Jp{}, the ladder, and the plaquette model. Fits for the ladder and
the plaquette model yield a common $\nu=0.709(6)$ while $\nu=0.689(5)$
is obtained for the \Jp{} model, in accordance with the previous
analysis.  Similarly the remaining exponents $\eta$ and $z$ are
determined to be $d-z-\eta=0.908(5)$ (\Jp) as well as
$d-z-\eta=0.971(2)$ (other models) with the obvious contrast. It is
easily checked, that the scaling law $2\beta=(d+z-2+\eta)\nu$ is
satisfied within error bars for all cases.  The exponent $\eta$ for
the \Jp{} model is thus given by $\eta=0.09(1)$, which is considerably
larger than the standard O(3) value.
\begin{figure}
\includegraphics[width=0.9\columnwidth]{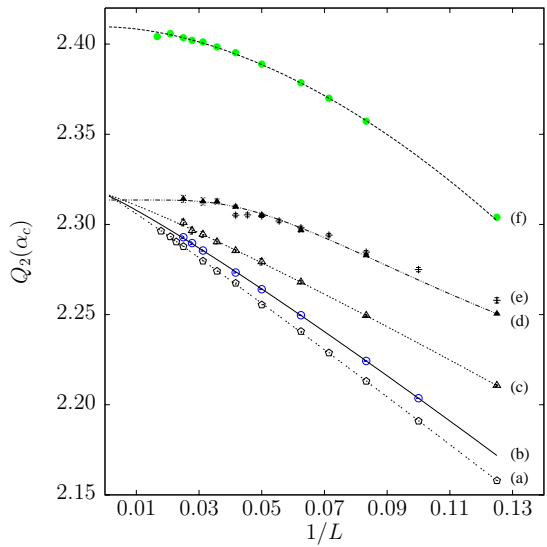}
\caption{\label{fig:Q2cmp} (color online). The critical Binder
  parameter in dependence on the lattice size $L$ for the (a)
  plaquette model, (b) ladder model, (c) bilayer, (d) Kondo lattice,
  (e) classical O(3) model, and (f) \Jp{} model. Data for the CaVO
  lattice is not shown as they overlap with curve (c).}
\end{figure}

Our findings are finally reinfored by comparing the Binder parameter
at the best known critical points for the models of
Table~\ref{tab:cmpQ2}. It is evident from Fig.~\ref{fig:Q2cmp} that all cases apart from the
\Jp{} model are in accordance with O(3) behavior. To make this
comparison even stronger we also include in Fig.~\ref{fig:Q2cmp} the value from Wolff cluster
simulations of the ordinary 3D classical Heisenberg model
\footnote{$Q_2$ is then also computed for the z-component of the
  magnetization and as an average over slices in z-direction. Note
  that $\langle (m^z)^4 \rangle/\langle (m^z)^2 \rangle^2$ is equal to
  $(9/5)\langle m^4\rangle/ \langle m^2 \rangle^2$.}.

\paragraph{Conclusion:}
In this Letter, we give comprehensive numerical evidence for an
unconventional universality class of the \Jp{} model based on data
collapsing analysis, scaling at criticality and by a comparison of the
Binder parameter for six different dimerized models. This shows that
there are nontrivial contributions to the quantum critical point
changing the critical exponents. Those contributions are triggered by
the special staggered arrangement of couplings. Our result challenges
the current understanding of quantum phase transitions in dimerized
quantum spin systems and it will be interesting to see which exact
theoretical mechanism accounts for the observed discrepancy.

\acknowledgments We acknowledge stimulating discussions with J.
Richter, D. Ihle, B. Berche, M. Hasenbusch, and S. Wessel.  We further
thank R. Kenna for corrections on the manuscript. S.W. is indebted for
support from the Studienstiftung des deutsches Volkes, the DFH-UFA, and
the graduate school ``BuildMoNa.'' This work was partially performed
on the JUMP computer of NIC at the Forschungszentrum J\"ulich under
project number HLZ12.

\bibliographystyle{apsrev}   
\bibliography{literature}

\end{document}